\documentclass[12pt]{article}
\usepackage[dvips]{graphics}
\usepackage{graphicx}
\usepackage{dcolumn}
\usepackage{bm}
\usepackage{amsmath}
\usepackage{amssymb}
\usepackage{mathptmx}
\usepackage{helvet}
\usepackage{courier}

\usepackage{makeidx}         % allows index generation
\usepackage{graphicx}        % standard LaTeX graphics tool

\usepackage{epsfig}
\usepackage{multicol}        % used for the two-column index
\usepackage[bottom]{footmisc}% places footnotes at page bottom
\usepackage{textcomp}
\usepackage{color}
\definecolor{darkblue}{rgb}{0,0,.5}
\definecolor{darkgreen}{rgb}{0,0.5,0}
\definecolor{darkred}{rgb}{0.5,0,0}
\usepackage[colorlinks = true,
            linkcolor = darkblue,
            urlcolor  = darkblue,
            citecolor = darkblue,
            anchorcolor = blue]{hyperref}

\newcommand{\beq}{\begin{equation}}
\newcommand{\eeq}{\end{equation}}
\newcommand{\beqa}{\begin{eqnarray}}
\newcommand{\eeqa}{\end{eqnarray}}
\newcommand{\bd}[1]{ \mbox{\boldmath $#1$}}

\begin{document}
\def\ii{\'\i}

\title{
Review on the pseudo-complex General Relativity and dark energy
}

\author{
Peter O. Hess$^{1,2}$ \\
{\small\it
$1$Instituto de Ciencias Nucleares, UNAM, Circuito Exterior,}\\
{\small\it
 C.U., A.P. 70-543, 04510, Mexico D.F., Mexico} \\ 
{\small\it
$^2$Frankfurt Institute for Advanced Studies, Wolfgang Goethe University,}\\
{\small\it 
Ruth-Moufang-Strasse 1, 60438 Frankfurt am Main, Germany}\\
}

%\date{\bf Version under development}

%\twocolumn[
%\begin{@twocolumnfalse}
\maketitle
\begin{abstract}
A review will be presented on the algebraic extension of the standard
Theory of Relativity (GR) to the pseudo-complex formulation (pc-GR). 
The pc-GR predicts the existence of a dark energy outside and inside the
mass distribution, corresponding to a modification of the GR- metric.
The structure of the emission profile of an accretion disc changes also inside
a star. Discussed are the consequences of the dark energy 
for cosmological models, permitting different outcomes on the evolution of 
the universe.
\end{abstract}
%\end{@twocolumnfalse}

\section{Introduction}
\label{secintro}

The Theory of General Relativity (GR) \cite{gravitation} is one of the best known theories tested
\cite{will}, mostly in solar system experiments. Also the loss of orbital energy in a binary system \cite{taylor}
was the first indirect proof for gravitational waves, which were finally detected in \cite{abbott}. In near future, the
{\it Event Horizon Telescope} \cite{EHT} will resolve the black hole in SgrA$^*$ and in the galaxy M87 and will 
publish their results soon..

Nevertheless, the limits of GR may be reached when strong gravitational fields are present, which can lead
to different interpretations of the sources of gravitational waves \cite{hess2016,hess2019}. 

A first proposal to extend GR was attempted by A. Einstein \cite{einstein1,einstein2} who introduced  complex values
metric $G_{\mu\nu}$ = $g_{\mu\nu}+iF_{\mu\nu}$, with $G_{\mu\nu}^* = G_{\nu\mu}$. The real part 
corresponds tio the standard metric, while the imaginary part defines the electromagnetic tensor. With this,
A. Einstein intended to unify GR with Electrodynamics. Another motivation to extend GR is published in \cite{born1,born2},
where M. Born investigated on how to recover the symmetry between
coordinates and momenta, which are symmetric in Quantum Mechanics but not in GR. To achieve his goal, he 
introduced also a complex metric, where the imaginary paert is momentum dependent. In \cite{caianello} this
was more elaborated, leading to the square of the length element ($c=1$)

\beqa
d\omega^2 & = & g_{\mu\nu} \left[ dx^\mu dx^\nu + l^2 du^\mu d u^\nu \right]
~~~,
\label{cai}
\eeqa
which implies maximal acceleration (see also \cite{book}). The ionteresting feature is that a minimal length "$l$" is
introduced as a {\it parameter} and Lorentz symmetry is, thus, automatically maintained, no deformation to
small lengths is necessary!

In \cite{kelly} the GR was algebraically extended to a series of variables and the solutions for the limit
of weak gravitational fields were investigated. As a conclusion, only real and pseudo-complex (called in \cite{kelly}
{\it hyper-complex}) make sense, because all others show either tachyon or ghost solutions, or both. Thus,
even the complex solutions don't make sense. This was the reason to concentrate on the pseudo-complex extension. 

In pc-GR, all the extended theories, mentioned in the last paragraph, are contained and the Einstein equations require an
energy-momentum tensor, related to vacuum fluctuations (dark energy), described by an asymmetric ideal fluid \cite{book}. 
Due to the lack of a microscopic theory, this dark energy is treated phenomenological. One possibility is to choose it
such that no event horizon appears, or barely still exists. The reason to do so is, that
in our philosophical understanding, {\it no theory should have a singularity}, 
even a coordinate singularity of the type of an event horizon encountered in a black hole. 
Though it is only a coordinate singularity,
the existence of an event horizon implies that even a black hole in a nearby corner cannot be accessed by an outside
observer. Its event horizon is a consequence of a strong gravitational field. 
Because no quantized theory of gravitation exists yet, 
we are lead to the construction of models for the distribution
of the dark energy.

In \cite{Nielsen2018} the pc-GR was compared to the observation of the amplitude for the inspiral process. As found,
the fall-off in $r$ of the dark energy has to be stronger than suggested in earlier publications. We will discuss this and what will
change, in the main
body of the text.

A general principle emerges, namely that
{\it mass not only curves the space} (which leads to the
standard GR) {\it but also changes the space- (vacuum-) structure in its
vicinity}, which in turn leads to an important deviation from the classical 
solution.

The consequences will be discussed in section
\ref{secappl}. There, also the cosmological effects are discussed, with different outcomes for the evolution of the dark
energy as function of time/radius of the universe. Another applications treats the interior of a stars, where
first attempts will be reported on how to stabilize a large mass. In section \ref{concl} Conclusions will be drawn.

\section{Pseudo-complex General Relativity (pc-GR)}
\label{secpcgr}

An algebraic extension of GR consists in a mapping of the real
coordinates to a different type, as for example 
complex or pseudo-complex (pc) variables

\beqa
X^\mu & = & x^\mu + I y^\mu
~~~,
\label{eq-1}
\eeqa
with $I^2 = \pm 1$ and 
where $x^\mu$ is the standard coordinate in space-time and
$y^\mu$ the complex component.
When $I^2 = -1$ it denotes {\it complex}
variables, while when $I^2 = +1$ it denotes pseudo-complex
(pc) variables. 
This algebraic mapping is just {\it one possibility} to explore extensions of GR.

In \cite{kelly} all possible extensions of real coordinates in GR where considered. It was found
that only the extension to pseudo-complex coordinated (called in \cite{kelly} {\it hyper-complex}) makes sense, 
because all others lead to tachyon and/or ghost solutions, in the limit of weak gravitational fields.

In what follows, some properties of pseudo-complex variables are resumed, which is important
to understand some of the consequences. 

\begin{itemize}

\item The variables can be expressed alternatively as

\beqa
X^\mu & = & X^\mu_+ \sigma_+ + X^\mu_- \sigma_-
\nonumber \\
\sigma_\pm & = & \frac{1}{2}\left( 1 \pm I\right)
~~~. 
\label{eq-2}
\eeqa

\item The $\sigma_\pm$ satisfy the relations

\beqa
\sigma_\pm^2 & = & \sigma_\pm ~,~ \sigma_+\sigma_-=0
~~~.
\label{eq-3}
\eeqa

\item Due to the last property in (\ref{eq-3}), when multiplied
one variable proportional to $\sigma_+$ by another one proportional
to $\sigma_-$, the result is zero, i.e. there is a 
{\it zero-divisor}. The variables, therefore, do not form
a {\it field} but a {\it ring}.

\item In both zero-divisor component ($\sigma_\pm$) the
analysis is very similar to the standard complex analysis.

\end{itemize}

In pc-GR the metric is also pseudo-complex

\beqa
g_{\mu\nu} & = & g^+_{\mu\nu}\sigma_+ + g^-_{\mu\nu}\sigma_-
~~~.
\label{eq-4}
\eeqa
Because $\sigma_+\sigma_-=0$ in each zero-divisor component one can construct independently a GR theory.

For a consistent theory, both zero-divisor components have to be connected! One possibility is
to define a modified variational principle, as done in \cite{hess2009}. Alternatively, one can implement
a constraint, namely that a particle should always move along a real path, i.e. that the 
pseudo-complex length element should be real.

The infinitesimal pc length element squared is given by (see also \cite{hess-ws})

\beqa
d\omega^2 & = & g_{\mu\nu} dX^\mu dX^\nu
\nonumber \\
& = & g^+_{\mu\nu} dX_+^\mu dX_+^\nu \sigma_+ + 
g^-_{\mu\nu} dX_-^\mu dX_-^\nu \sigma_-
\label{eq-5}
\eeqa
as written in the zero-divisor components. In terms of
the pseudo-real and pseudo-imaginary components, we have

\beqa
d\omega^2 & = & 
g^s_{\mu\nu} \left( dx^\mu dx^\nu 
+ dy^\mu dy^\nu \right) 
+ g^a_{\mu\nu} \left( dx^\mu dy ^\nu + dy^\mu dx^\nu \right)
\nonumber \\
&& + 
I\left[
g^a_{\mu\nu} \left( dx^\mu dx^\nu 
+ dy^\mu dy^\nu \right) 
+ g^s_{\mu\nu} \left( dx^\mu dy ^\nu + dy^\mu dx^\nu \right)
\right]
~~~,
\label{eq-6a}
\eeqa
with
$g^s_{\mu\nu}$  =  $\frac{1}{2}\left( g^+_{\mu\mu}
+ g^-_{\mu\nu} \right)$ and 
$g^a_{\mu\nu}$ =  $\frac{1}{2}\left( g^+_{\mu\mu}
- g^-_{\mu\nu} \right)$.
The upper indices $s$ and $a$ refer to a {\it symmetric}
and {\it antisymmetric} combination of the metrics. For the case when
$g_{\mu\nu}=g_{\mu\nu}^+ = g_{\mu\nu}^-$, i.e.$ g_{\mu\nu}^a=0$, leads to

\beqa
g_{\mu\nu} \left( dx^\mu dx^\nu 
+ dy^\mu dy^\nu \right) 
+ 
I g_{\mu\nu} \left( dx^\mu dy ^\nu + dy^\mu dx^\nu \right)
~~~.
\label{eq-6b}
\eeqa
Identifying $y^\mu = l u^\mu$, where $l$ is an infinitesimal length and $u^\mu$ the
4-velocity, one obatines the length element defined in \cite{caianello}.
It also contains the line element as proposed in \cite{born1,born2}, where the $y^\mu$ is proportional 
to the momentum component $p^\mu$ of a particle.
 However, this identification of
$y^\mu$ is only valid in a flat space, where the second term in (\ref{eq-6b}) is just the scalar
product of the 4-velocity ($u^\mu = \frac{dx^\mu}{d\tau}$ to the 4-acceleration
($y^\mu = \frac{d^2 x^\mu}{d\tau^2}$).

The connection between the two zero-divisor components 
is achieved, requiring that the infinitesimal length element
squared in (\ref{eq-6a}) {\it is real}, i.e., in terms of
the $\sigma_\pm$ components it is

\beqa
\left(\sigma_+-\sigma_-\right) 
\left(
g^+_{\mu\nu} dX_+^\mu dX_+^\nu
-
g^-_{\mu\nu} dX_-^\mu dX_-^\nu
\right) & = & 0
~~~.
\label{eq-7a}
\eeqa

Using the standard variational principle with a Lagrange multiplier, to account for the constraint, leads
to an
additional contribution in the Einstein equations, interpreted
as an energy-momentum tensor.

The action of the pc-GR is given by \cite{hess-ws}

\beqa
S=\int dx^4 \sqrt{-g}\left({\cal R}+2\alpha \right)
~~~,
\label{action}
\eeqa
where ${\cal R}$ is the Riemann scalar.
The last term in the action integral allows to introduce the cosmological constant
in cosmological models, where $\alpha$ has to be constant in order not to violate the
Lorentz symmetry. This changes when a system with
a uniquely defined center is considered, which has spherical 
(Schwarzschild) or axial (Kerr) symmetry. 
In these cases, the
$\alpha$ is allowed to be a function in $r$, for the Schwarzschild solution, and a 
function in $r$ and $\vartheta$, for the Kerr solution.  

The variation of the action with respect
to the metric $g_{\mu\nu}^\pm$ leads to the equations of motion

\beqa
{\cal R}^\pm_{\mu\nu} - \frac{1}{2}g^\pm_{\mu\nu}{\cal R}_\pm
& = &
8\pi T_{\pm~\mu\nu}^\Lambda
\nonumber \\
{\rm with} &&
\nonumber \\
 T_{\pm~\mu\nu}^\Lambda & = & \lambda u_\mu u_\nu + \lambda \left( {\dot y}_\mu {\dot y}_\nu \pm u_\mu {\dot y}_\nu
\pm u_\nu {\dot y}_\mu \right) +\alpha g_{\mu\nu}^\pm
~~~,
\label{II-X-eq-8}
\eeqa 
in the zero-divisor component, denoted by the independent unit-elements $\sigma_\pm$.
These equations still contain the effects of a minimal length parameter $i$, as shown in \cite{caianello}.
because the effects of a minimal length scale is difficult to measure, maybe not possible at all, we neglect it,
which corresponds to map the above equations to their real part, giving

\beqa
{\cal R}_{\mu\nu} - \frac{1}{2}g^\pm_{\mu\nu}{\cal R}
& = &
8\pi T_{\mu\nu}^\Lambda
~~~.
\label{II-X-eq-8a}
\eeqa 

The
$T^\Lambda_{\mu\nu}$, ${\cal R}^\pm_{\mu\nu}$ is real and is now
given by
\cite{hess-ws}

\begin{equation}
   T^\Lambda_{\mu \nu , R}=(\varrho_\Lambda+
	p^\Lambda_\vartheta)u_\mu u_\nu+
	p^\Lambda_\vartheta g_{\mu \nu}+
	\left( p^\Lambda_r-p^\Lambda_\vartheta\right)k_\mu k_\nu~~~,
\label{eq-tmunu}
\end{equation}
where $p^\Lambda_\vartheta$ and radial $p^\Lambda_r$ are
the tangential and pressure respectively.
For an isotropic fluid we have $p^\Lambda_\vartheta$ = $p^\Lambda_r$
=$ p^\Lambda$.  
The $u^\mu$ are the components of the 4-velocity of the
elements of the fluid and $k^\mu$ is a space-like vector
($k_\mu k^\mu = 1$) in the radial direction. It satisfies
the relation $u_\mu k^\mu = 0$.
The fluid is anisotropic due to the presence of $y_\mu$.
The $\lambda$ and $\alpha$ are
related to the pressures as \cite{hess-ws}

\beqa
\lambda & = & 8\pi {\tilde \lambda} ~,~ \alpha ~=~ 8\pi {\tilde \alpha}
\nonumber \\
{\tilde \lambda } & = & \left( p^\Lambda_\vartheta + \varrho_\Lambda \right) ~,~
{\tilde \alpha} ~=~ p^\Lambda_\vartheta ~,~
{\tilde \lambda} y_\mu y_\nu~=~ \left( p^\Lambda_r - p^\Lambda_\vartheta \right) k_\mu k_\nu
~~~.
\label{parameters}
\eeqa

The reason, why the dark energy outside a mass distribution has to be an anisotropic fluid,
is understood contemplating the
{\it Tolman-Oppen\-heimer-Volkov} (TOV) equations \cite{adler}
for an isotropic fluid: The TOV equations 
relates the derivative of the dark-energy pressure with respect to $r$ 
(for an isotropic fluid, the tangential pressure  has to be the same as the radial pressure, i.e., 
$p^\Lambda_\vartheta = p^\Lambda_r = p^\Lambda$ )
to the dark energy density $\varrho_\Lambda$.
Assuming the isotropic fluid and 
that the equation of state for the dark energy is
$p^\Lambda = -\varrho_\Lambda$, the factor
$\left( p^\Lambda + \varrho_\Lambda \right)$ in the TOV
equation for $\frac{dp^\Lambda}{dr}$ is zero, i.e.,
the pressure derivative is zero. As a result the
pressure is constant and with the equation of state
also the density is constant, which leads to a contradiction. 
Thus, the fluid has to be anisotropic, due to an additional term, allowing the pressure to to fall off
as a function on increasing distance.
The additional term in the radial pressure $\frac{dp^\Lambda_r}{dr}$, 
added to the TOV equation, is given by
$\frac{2}{r}\Delta p^\Lambda$ = 
$\frac{2}{r}\left(p^{\Lambda}_{\vartheta}-p^{\Lambda}_{r}\right)$
\cite{rodriquez2014a}. 
	
For the density one has to apply a phenomenological model, due to the lack of
a quantized theory of gravity. What helps is to recall one-loop calculations in
gravity \cite{birrell}, where vacuum fluctuations result due to the non-zero back ground curvature
(Casimir effect).
Results are presented in \cite{visser-boul}, where at large distances the density falls
off approximately as $1/r^6$. The semi-classical
Quantum Mechanics \cite{birrell} was applied, which assumes a {\it fixed}
back-ground metric and is thus only valid for weak gravitational fields (weak compared to the solar system).
Near the Schwarzschild radius 
the field is very strong which is exhibited by a singularity in the energy density, which is proportional to
$\frac{1}{\left( 1- \frac{2m}{r}\right)^2}$, with $m$ a constant mass parameter \cite{visser-boul}.

Because we treat the vacuum fluctuations as a classical 
ideal anisotropic fluid, we are free to propose 
a different fall-off of the 
negative energy density, which is finite at the Schwarzschild
radius. In earlier publications the density did fall-off proportional to $1/r^5$. However, in \cite{Nielsen2018}
it is shown that this fall-off has to be stronger. Thus, in this contribution we will also discuss a variety of fall-offs as a funtion
of a parameter $n$, i.e., proportional to $\frac{B_n}{r^n}$, where $B_n$ describes the coupling of the dark energy
to the mass. 

With the assumed density, the metric for the Kerr solution changes to
\cite{schoenenbach2012,Schoenenbach2014c}

\beqa
g_{00} & = & -\frac{ r^2 - 2m r  + a^2 \cos^2 \vartheta + \frac{B_n}{(n-1)(n-2)r^{n-2}} }{r^2 + a^2 \cos^2\vartheta}~~~, \nonumber \\
g_{11} & = & \frac{r^2 + a^2 \cos^2 \vartheta}{r^2 - 2m r + a^2 +  \frac{B_n}{(n-1)(n-2)r^{n-2}}  }~~~, \nonumber \\
g_{22} & = &  r^2 + a^2 \cos^2 \vartheta~~~,  \nonumber \\
g_{33} & = & (r^2 +a^2 )\sin^2 \vartheta + \frac{a^2 \sin^4\vartheta \left(2m r -  \frac{B_n}{(n-1)(n-2)r^{n-2}}  \right)}{r^2 + a^2 \cos^2 \vartheta}~~~,  \nonumber \\
g_{03} & = & \frac{-a \sin^2 \vartheta ~ 2m r + a \frac{B_n}{(n-1)(n-2)r^{n-2}}   \sin^2 \vartheta }{r^2 + a^2 \cos^2\vartheta}~~~,
\label{V-eq:kerrpseudo}
\eeqa
where $0\leq a \leq m$ is the spin parameter of the Kerr 
solution and $n$ = 3, 4, .... For $n=2$ the old ansatz is achieved.
The Schwarzschild solution is obtained, setting $a=0$.
The parameter $B_n=b_n m^n$ measures the coupling of the
dark energy to the central mass. The definition of $n$ here is related to the $n_N$ in \cite{Nielsen2018} by
$n_N = n-1$.

When no event horizon is demanded, the  parameter $B_n$ has a
lower limit given by 

\beqa
B_n & > & \frac{2(n-1)(n-2)}{n}\left[\frac{2(n-1)}{n}\right]^{n-1}~{\rm m}^n ~=~b_{\rm max} {\rm m}^n
~~~.
\label{B}
\eeqa
For the equal sign, an event horizon is located at

\beqa
r_h & = & \frac{2(n-1)}{n}~{\rm m}
~~~,
\label{horizon}
\eeqa
e.g.., $\frac{4}{3}$ for $n=3$ and $\frac{3}{2}$ for $n=4$. 

\section{Applications}
\label{secappl}

\subsection{Motion of a particle in a circular orbit}
\label{seccircular}

In \cite{MNRAS2013} the motion of a particle in a circular orbit was investigated. This section were first discussed
in \cite{MNRAS-I,MNRAS-II}.

\begin{figure}
\centerline{
\rotatebox{180}{\resizebox{220pt}{220pt}{\includegraphics[width=0.60\textwidth,angle=90]{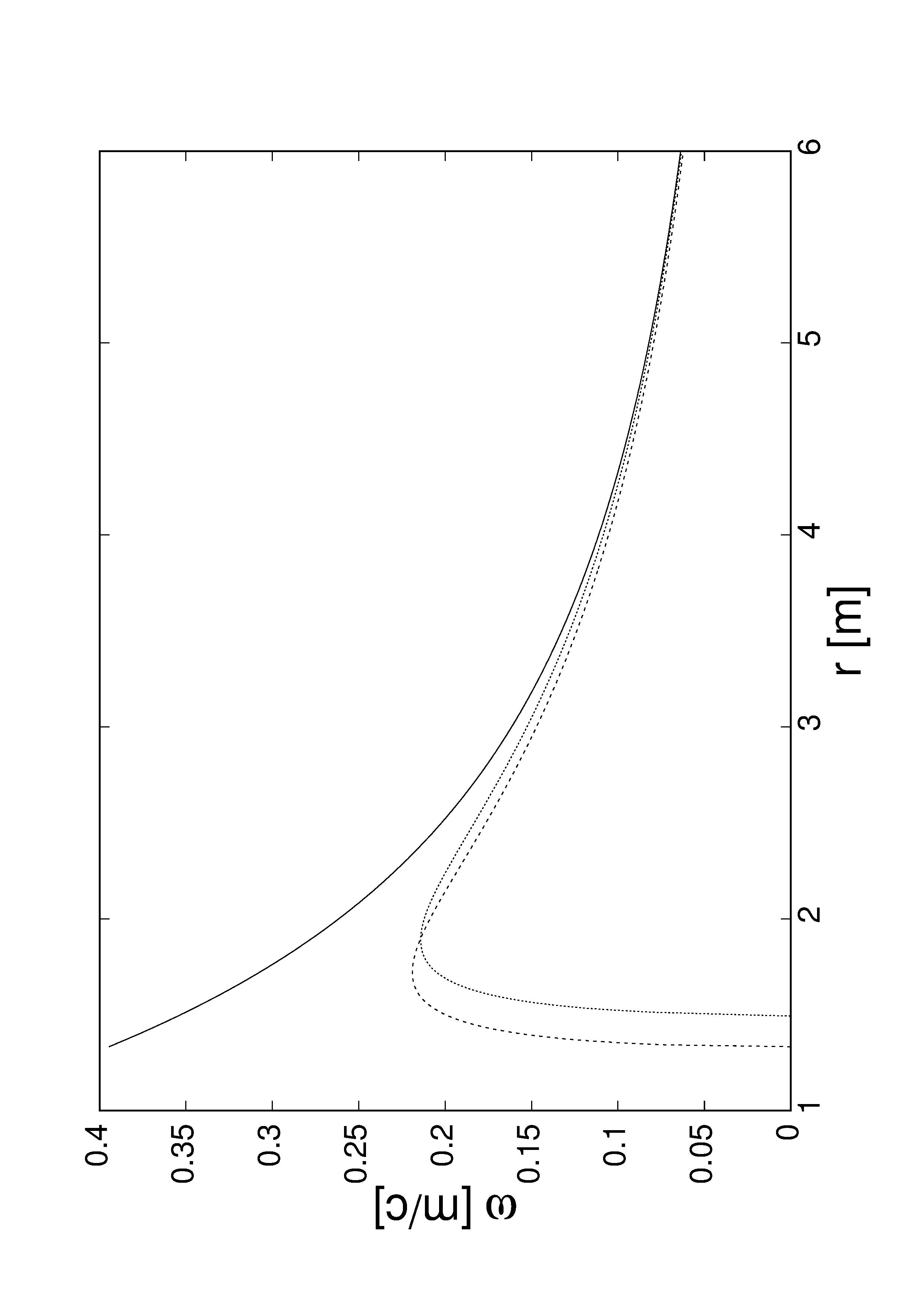}}}}
\caption{ The orbital frequency of a particle in  a circular orbit for the case GR (upper curve) and for $n=3$ (dashed curve) 
and $n=4$ (dotted curve) \cite{MNRAS-II,MNRAS-II}. \label{omega}
}
\end{figure}

\begin{figure}[ht]
\begin{center}
\rotatebox{270}{\resizebox{230pt}{300pt}{\includegraphics[width=0.5\textwidth]{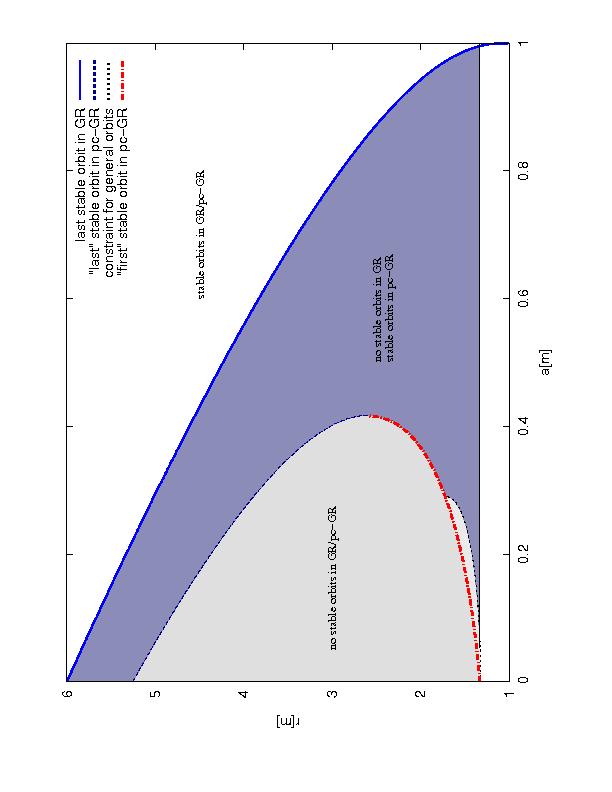}}}
\end{center}
\caption{
The position of the {\it Innermost Stable Circular Orbit} (ISCO) 
is plotted versus the rotational 
parameter $a$. The upper curve corresponds to GR and the 
lower curves to pc-GR. The light gray shaded region corresponds to
a forbidden area for circular orbits within pc-GR. For small
values of $a$ the ISCO in pc-GR follows more or less the
one of GR, but at smaller values of $r$. From a certain $a$ on stable orbits
are allowed until to the surface of the star (for $n=3$ this limit is approximately 
0.4, calculations for $n=4$ are in progress).
\label{stable}}
\end{figure}

The main results is resumed in the Figures 
\ref{omega} and \ref{stable}. In Fig. \ref{omega} the orbital frequency, in units of $\frac{c}{m}$, is depicted versus the radial
distance $r$, in units of $m$, for a rotational parameter of $0.995m$. The function for the orbital frequency, in prograde orbits,
is given by

\beqa
\omega_n & = & \frac{1}{a+\sqrt{\frac{2r}{h_n(r)}}}
\nonumber \\
h_n(r) & = & \frac{2}{r^2} - \frac{nB_n}{(n-1)(n-2)r^{n+1}}
~~~.
\label{omegapro}
\eeqa

The upper curve in Fig. \ref{omega} corresponds to GR while the two  lower ones to pc-GR with $n=3$ (dashed curve)
and $n=4$ (dottet curve).
The curve shows a maximum at 

\beqa
r_{\omega_{\rm max}} & = & \left[ \frac{n(n+2)b_{\rm max}}{6(n-1)(n-2)}\right]^{\frac{1}{n-1}}~{\rm m}
~~~,
\label{rom}
\eeqa
which, for $b_n=b_{\rm max}$ as given in (\ref{B}), is independent of the value of $a$,
after which it falls off toward the
center and reaches zero at $r_h$ (Eq. (\ref{horizon})).

\beqa
r_h=\frac{2(n-1)}{n}~{\rm m}
~~~,
\label{romegamax}
\eeqa
which is {\it independent} on the rotational parameter $a$.
After the maximum the curve falls off toward smaller $r$. These features will be important for the understanding of the
emission structure of an accretion disc (see next subsection).

As one can see, the difference between $n=3$ and $n=4$ is minimal and, thus, will not change the qualitative results
as obtaioned for $n=3$ in former publications. The position of the maximum, which gives the position of the dark 
ring discussed below, is approximately the same in both cases. For  $b_n \rightarrow 0$ the curve approaches the one
for GR.

In Fig. \ref{stable} the last stable orbit, for $n=3$, is plotted versus the rotational parameter $a$. The solid enveloping curve is the
result for GR. For $a=0$ the last stable orbit in GR is at $6m$, while in pc-GR it is further in. The dark gray 
shaded area describe stable
orbits in pc-GR and thelight  gray area unstable orbits. 
The pc-GR follows closely the GR with a greater deviation for larger $a$. At about $a=0.45 m$
(for $n=3$, for $n=4$ its value is a litle bit larger) all orbits in pc-GR are stable 
up to the surface of the star, which is estimated to ly at approximately $\frac{4}{3}m$. For $a=m$, in GR the last stable orbit is at
$r=m$.

\subsection{Accretion discs}
\label{subsecdisc}

In order to connect to actual  .observations \cite{EHT},
one possibility is to simulate accretion
discs around massive objects as the one in the center of
the elliptical galaxy M87. The underlying theory was published
by D. N. Page and K. S. Thorne \cite{pagethorne} 
in 1974. The basic
assumptions are (see also \cite{MNRAS2014})

\begin{itemize}

\item A thin, infinitely extended accretion disc. This is a simplifying assumption. A real accretion disc can be a torus.
Nevertheless, the structure in the emission profile will be similar, as discussed here. This discs are easier to calculate 

\item An energy-momentum tensor is proposed which includes
all main ingredients, as mass and electromagnetic contributions.

\item Conservation laws (energy, angular momentum and mass) 
are imposed in order to obtain the flux function, the  main  result of \cite{pagethorne}. 

\item The internal energy of the disc is liberated via shears 
of neighboring orbitals and distributed from orbitals of 
higher frequency to those of lower frequency.

\end{itemize}

How to deduce finally the flux is described in detail in \cite{hess-ws}.

\begin{figure}[t]
\rotatebox{0}{\resizebox{230pt}{230pt}{\includegraphics[width=0.5\textwidth]{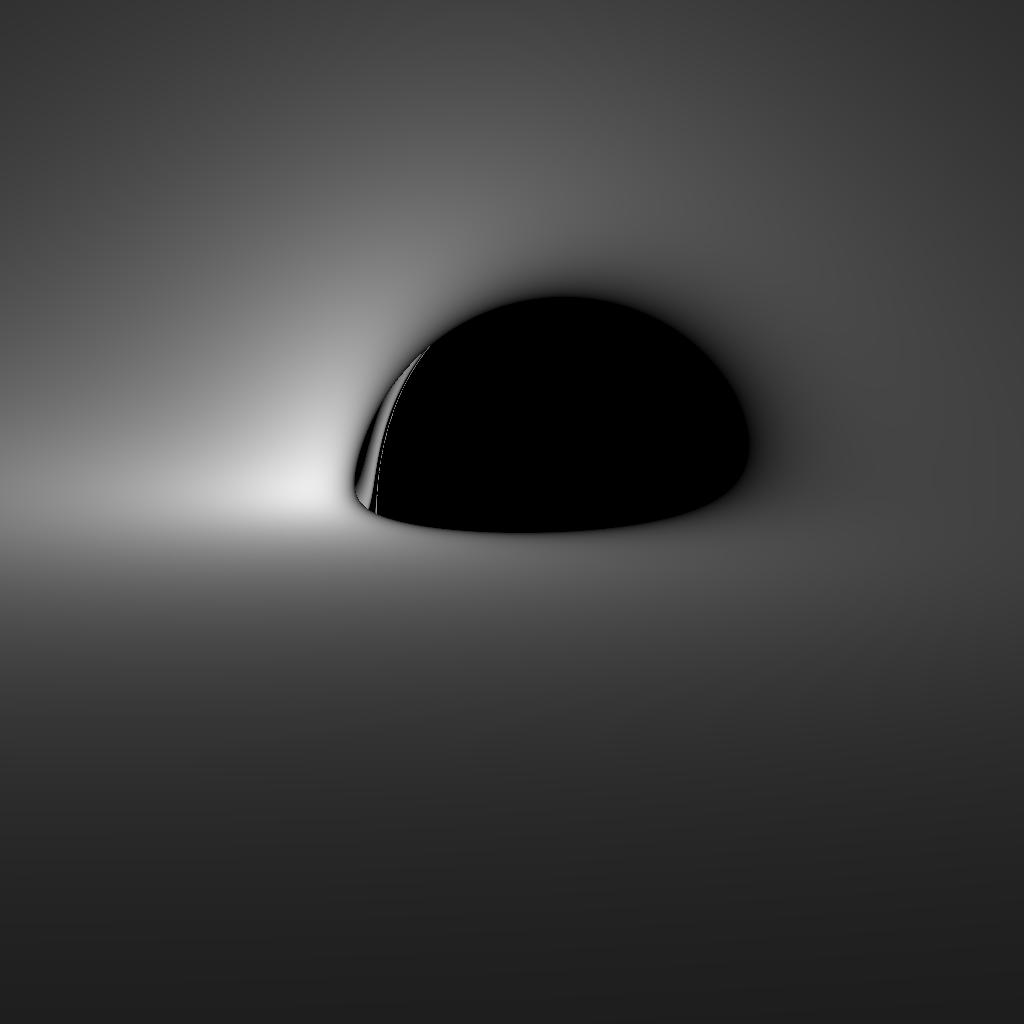}}}
\rotatebox{0}{\resizebox{230pt}{230pt}{\includegraphics[width=0.5\textwidth]{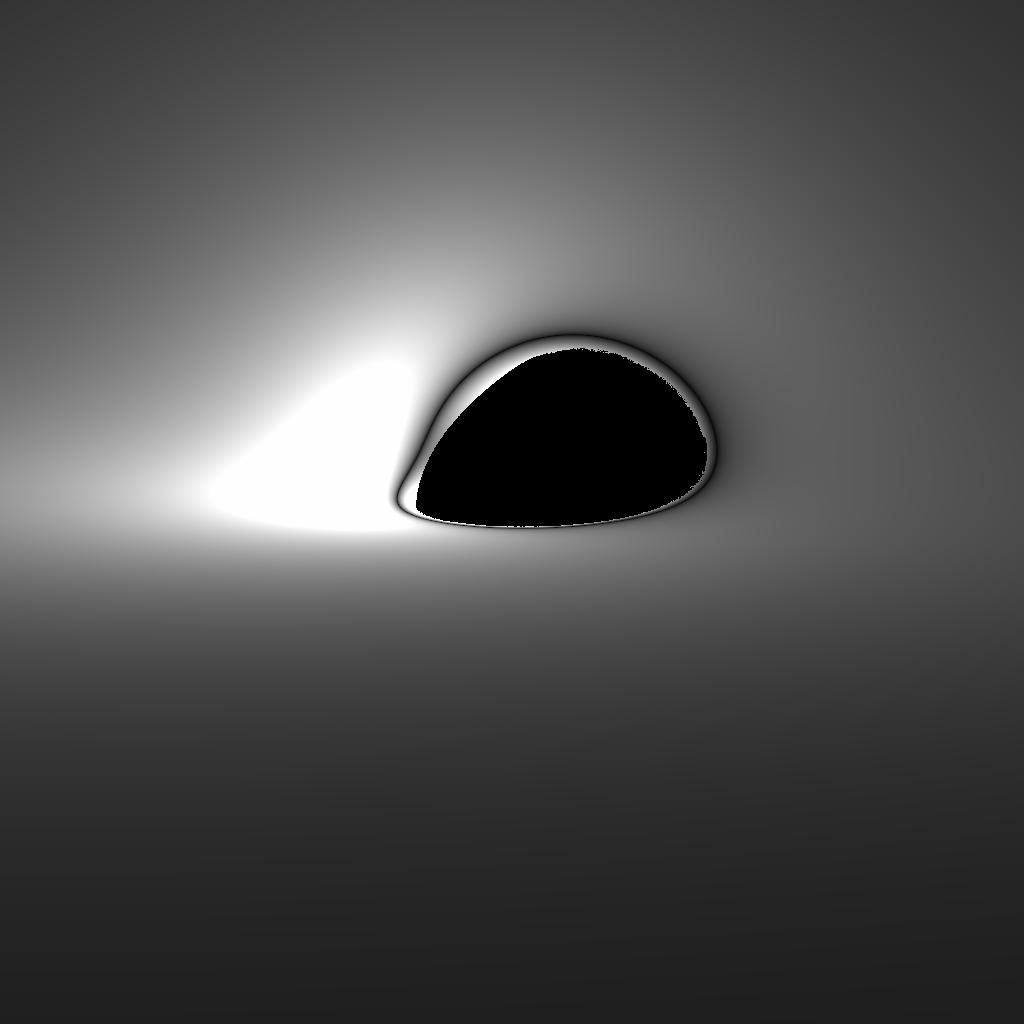}}}
\caption{ 
Infinite, counter clockwise rotating geometrically thin
accretion disc around static and rotating compact
objects viewed from an inclination of 80$^o$.
The left panel shows the disc model by
\cite{pagethorne} in pc-Gr, with $a=0$. The right panel shows the
modified model, including pc-GR correction terms as
described in the text. \label{discs}
}
\end{figure}

In order to understand within pc-GR the structure of 
the emission profile in the accretion disc, we have to get back to the discussion in the last subsection.
The local heating of the accretion disc is determined by the gradient of orbital frequency, when going further
inward (or outward). At the maximum, neighboring orbitals have
 nearly the same orbital frequency, thus, friction is low. On the other hand, above and especially below the position of the
maximum the change in orbital frequency is large and the disc gets heated. At the maximum the heating is minimal which will
be noticeable by a dark ring. Further inside, the heating increases again and a bright ring is produced.

The above consideration is relevant for $a$
larger than approximately 0.4, as can be seen from Figure 
\ref{stable} (for explanations, see the figure caption) and
\cite{MNRAS2013}. For lower values of $a$, in pc-GR the last stable orbit follows the one of GR, but with lower values 
for the position of the ISCO.
As a consequence, the particles reach further inside and due to the decrease of the potential, more energy is released, producing a brighter disc. However, the last stable orbit
in pc-GR does not reach $r_{\omega_{{\rm max}}}$. 
This changes when $a$ is a bit larger than 0.4. Now, 
$r_{\omega_{{\rm max}}}$ is crossed and the existence of the maximum of $\omega$ has to be taken into account as explained above. 

Some simulations are presented in Figure \ref{discs}. The 
line of sight of the observer to the accretion disc is 
80$^o$ (near to the edge of the accretion disc), 
where the angle refers to the one between the axis
of rotation and the line of sight. Two rotation parameters
of the Kerr solution are plotted, namely $a=0$ (no rotation of the star, corresponding to the Schwarzschild solution) and
nearly the maximal rotation $a=0.9$~m.

As a global feature, the accretion  disc in pc-GR
appears brighter, which is due to the fact that the
disc reaches further inside where the potential is deeper, thus releasing more
gravitational energy, which is then distributed within the disc.

The reason for the dark fringe and bright ring was explained above due to the variability of the friction.
The dark ring is the position of the maximum
of the orbital frequency. An observed position of a dark ring can, thus, be used to determine $n$.

The differences in the structure of an accretion disc gives us a clear observational criteria to
distinguish between GR and pc-GR. Though, we used here a simple model and an accretion disc may be better described by 
a torus, the discussed structures remain!  Unfortunately, this is for the moment
the only clear prediction to differentiate pc-GR from GR. In the next subsection we will discuss gravitational waves and
we will see pc-GR and GR give different interpretations of the source, though, the final outcome is the same.

\subsection{Gravitational waves in pc-GR}
\label{subsecgravwaves}

In \cite{abbott} the first observed gravitational wave event
was reported. In \cite{hess2016} this gravitational event was investigated within the pc-GR.

Using GR and the mass-point approximation
for the two black holes, before the merging, a relation is
obtained between
the observed frequency and its temporal change to the chirping mass
${\cal M}_c$,
namely \cite{Maggiore2008}

\beqa
{\cal M}_c ~=~
{\widetilde {\cal M}}_c F_{\omega}(r)
& = &
\frac{c^3}{G}\left[ \frac{5}{96\pi^\frac{8}{3}} \frac{df_{{\rm gw}}}{dt} 
f_{{\rm gw}}^{-\frac{11}{3}}
\right]^{\frac{3}{5}}
~~~.
\label{eq1}
\eeqa
substituting on the right hand side the observed frequency and its change and using
$F_{\omega}(r)=1$ for GR, the interpretation of the source of the gravitational waves  is of two black holes of about
30 solar masses each which fuse to a larger one of less than 60 solar masses. The difference in energy is radiated 
away as gravitational waves.
However, these changes in pc-GR, where the two
black holes can come very near to each other. Unfortunately, the point mass approximation is not applicable, though
in \cite{hess2016} this approximation was still used in order to show in which direction the interpretation of the
source changes. In pc-GR ($n=3$) $F_{\omega}(r)$ = $\left[1-\frac{3b}{4}\left(\frac{m}{r}\right)^2\right]$, which for $b_n$
given by the right hand side of Eq. (\ref{B})
is exactly zero. Therefore, a range of the last possible distances of the two black holes before
merging was assumed. On the left hand side of Eq. (\ref{eq1}), the function$ F_{\omega}(r)$ becomes very small near where
the two in-spiraling black holes merge. Thus, the chirping mass ${\widetilde {\cal M}}_c$ must be much larger than
the chirping mass ${\cal M}$ deduced in GR.  

The main result is that the source in pc-GR corresponds to two black holes with several
thousand solar masses. This may be related to the merger of two primordial galaxies whose central black hole
subsequently merges. One way to distinguish the two predictions is to look for light events very far way. If for observed
gravitational wave events in future, there is a consistent appearance of light events much father away as the distance 
deduced from GR, then this might   be in favor for pc-GR. However, all the prediction depends on the assumption that
the point mass approximation is still more or less valid when the two black holes are near together, which is not very good!
In \cite{Nielsen2018} the inspiral frequency was determined within pc-GR, for various values of $n$, which is rekated
to gthe one used in \cite{Nielsen2018} by $n_N=n-1$. As demonstrated, the wave form cannot be 
reproduced satisfactorily for $n=3$, thus, it has to be increased and let us to investigate the dependence of
the results as a function in $n$.

\begin{figure}[t]
\rotatebox{0}{\resizebox{600pt}{600pt}{\includegraphics[width=0.5\textwidth]{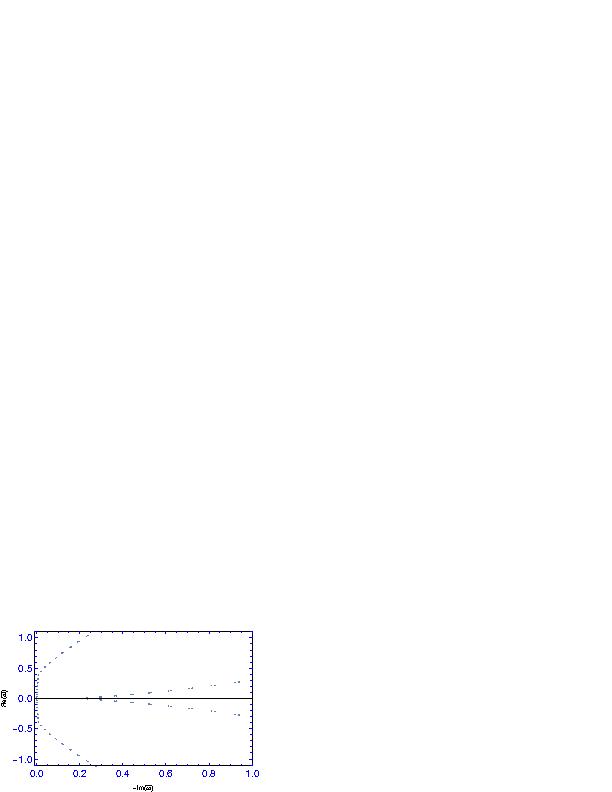}}}
\caption{ 
Axial gravitational modes inpc-GR. The vertical axis gives the real part of ${\widetilde \omega}= m \omega$ 
while the horizontal axis depicts the negative of its imaginary part.\label{axial}}
\end{figure}

In \cite{hess2019}  the Schwarzschild case was considered and the Regge-Wheeler, for negative parity solutions, \cite{regge}
and Zerrilli equations, for positive parity solutions, \cite{chandra} were solved, using an iteration method \cite{cho}. 
Due to a symmetry, in GR the two type of solutions have {\it the same} frequency spectrum \cite{chandra}, which unfortunately is
lost in pc-GR. For pc-GR, the spectrum of frequencies for {\it axial modes} show a convergent behavior for the frequencies,
which is shown in Fig. \ref{axial}. A negative imaginary part indicates a stable mode, which turns out to be the case.
For an increasing imaginary part the convergence is less sure. Unfortunately,
for the polar modes no convergence for the polar modes were obtained up to now.

Another problem is to distinguish between GR and pc-GR. It depends very much on the observation of the ring-down
frequency of the merger \cite{hess2019}, which is not very well measured yet. Without it, we are not able to distinguish between
both theories and various possible scenarios can be obtained in pc-GR \cite{hess2019,book}.

\subsection{Dark energy in the universe}
\label{cosm}

\begin{figure}[t]
\rotatebox{0}{\resizebox{250pt}{250pt}{\includegraphics[width=0.5\textwidth]{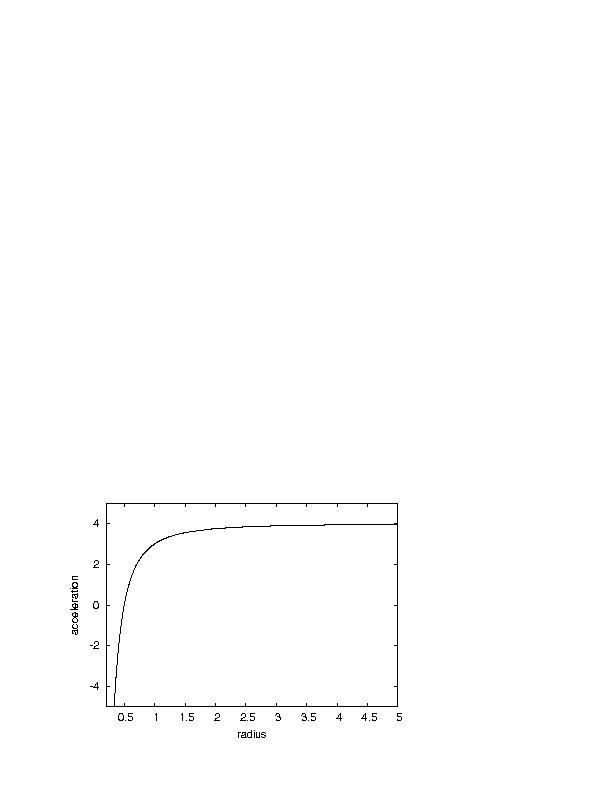}}}
\rotatebox{0}{\resizebox{250pt}{250pt}{\includegraphics[width=0.5\textwidth]{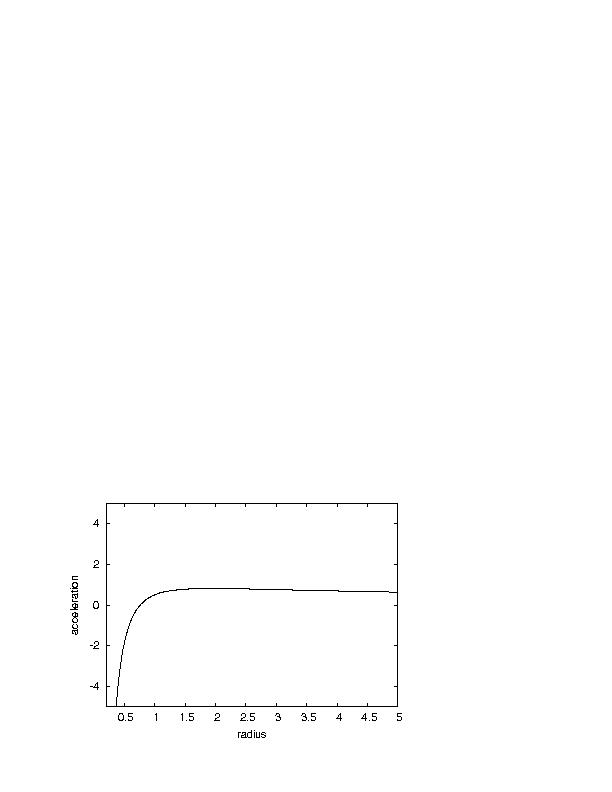}}}
\caption{ 
The left panel shows a case where the acceleration of the universe approaches a constant value and the right panel
a case where the acceleration slowly approaches zero for $t \rightarrow \infty$. The figures are taken from
\cite{book,leila}. 
 \label{RW}}
\end{figure}

The pc-Robertson-Walker model is presented in detail in \cite{book,leila}. The main results will be resumed in this subsection.

The line element in gaussian coordinates has the form

\beqa
d\omega^2 & = & (dt)^2 - a(t)^2  \frac{1}
{\left( 1+\frac{ka(t)^2}{4a_0^2} \right)^2} \left( dR^2 + R^2 d\vartheta^2
+ a(t)^2 {\rm sin}^2\vartheta d\varphi^2 \right)
~~~,
\label{rel-2}
\eeqa
where $R$ is the radius of the universe and $k$ is a parameter and the energy density of matter was 
assumed to be homogeneous. The value $k=0$ corresponds to a flat universe, which will be taken here.

The corresponding Einstein equations were solved and an equation for the radius $a(t)$ if the universe was 
deduced \cite{book}:

\beqa
{\bd a(t)}^{\prime\prime} & = & \frac{4\pi G}{3} (3\beta - 1)\Lambda
a(t)^{3(\beta - 1)+1} \nonumber \\
&& - \frac{4\pi G}{3} (1+3\alpha) \varepsilon_{0}
{\bd a(t)}^{-3(1+\alpha)+1}
~~~,
\label{rediusU}
\eeqa
where $G$ is the gravitational constant and $\beta$, $\Lambda$ are parameters of the theory. The equation of state is
set as $p = \alpha \varepsilon$, where $\varepsilon$ is the matter density and $\alpha$ is set to zero for {\it dust}.  

Two particular solutions are shown in Fig. \ref{RW}. Shown is the acceleration of the universe as
a function  of the radius $a(t)$.
The left panel shows the result for $\beta=\frac{1}{2}$ and
$\Lambda=3$ and on the right hand side the parameters $\beta$ and $\Lambda$ are set to $\frac{2}{3}$ and 4,
respectively. The left figure correspond to a solution where the acceleration tents to a constant, i.e., the universe
will expand for ever with an increasing acceleration, while in the right figure the acceleration 
tends slowly to zero for very large $a(t)$. In both
examples the universe expands for ever. These are not the only solutions, also one where the universe collapses again
is possible. 

These results are not very predictive, because one can obtain several possible outcomes, depending on the values
of $\beta$ and $\Lambda$. Nevertheless, they show that possible scenarios for the future of our
universe are still possible.

\subsection{Interior of stars}
\label{subsecneutron}

For the description of the interior of a star one needs the equation of state of matter
and the coupling of the dark-energy with the matter.
For the equation of state one can use the model presented in \cite{dexheimer}, which also takes into account nuclear
and meson resonances. However, these approximations will loose their validity when the matter density is too large.
the situation is worse for the dark-energy contribution and it is twofold: i) one has to know how the dark-energy evolves within
the star (presence of matter) and ii) how it is coupled to the matter itself. Both are not known and we have to rest on incomplete models. Alternatively, one can approach the problem with a very interesting and distinct model to simulate the dark energy, 
as done in \cite{volkmer1,volkmer2,volkmer3,volkmer4}, where
compact and dense objects were investigated within the pc-GR and maximal masses were also deduced.

In \cite{rodriquez2014a} a simple coupling model of dark-energy to the mass density was proposed

\beqa
\epsilon_\Lambda & = & \alpha \rho_m
~~~,
\label{rod1}
\eeqa
where the index $\Lambda$ refers to the dark energy and $\rho_m$ to the mass density. 
In this proposal the dark energy follows
neatly the mass distribution. The Tolman-Oppenheimer-Volkoff (TOV) equations have to be solved, which is doubled in 
number,  one treating the mass part and the other the dark-energy part (for more details see \cite{book} and 
\cite{rodriquez2014a}).

\begin{figure}[t]
\begin{center}
\rotatebox{0}{\resizebox{500pt}{500pt}{\includegraphics[width=0.5\textwidth]{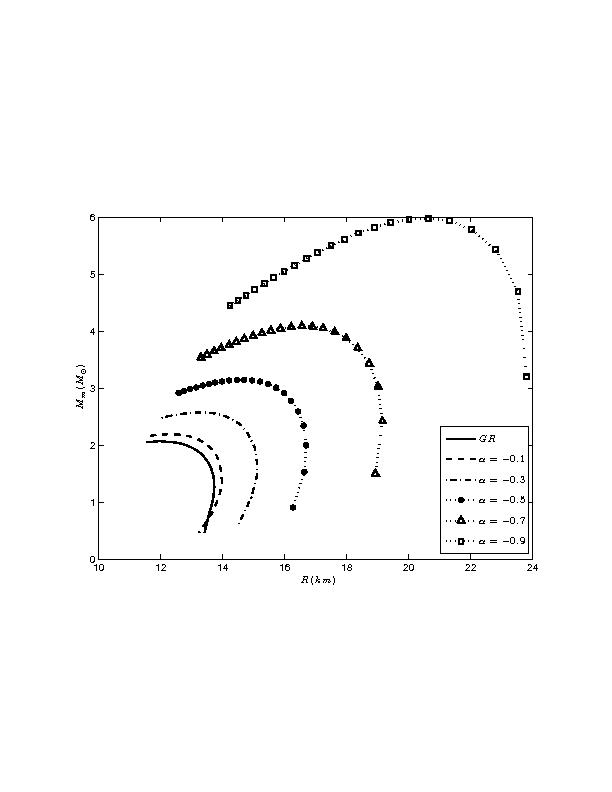}}}
\end{center}
\caption{ 
The figure shows the
dependence of the mass of the  star as a function of its radius
(figure taken from \cite{rodriquez2014a}). \label{neutron-isaac}}
\end{figure}

A particular result is shown in Fig. \ref{neutron-isaac}, showing the mass of the star versus its radius.
Curves are depicted for various values of the proportionality factor $\alpha$. As can be seen, the model can reproduce stable stars up to 6 solar masses, which shows that the dark-energy stabilizes stars with larger masses.
However, no stars with larger masses can be constructed, because for larger values of $\alpha$ 
and/or larger masses the repulsion due
to the dark energy becomes too large near the surface and outer surface layers are evaporated.

\begin{figure}[t]
\rotatebox{0}{\resizebox{600pt}{600pt}{\includegraphics[width=0.5\textwidth]{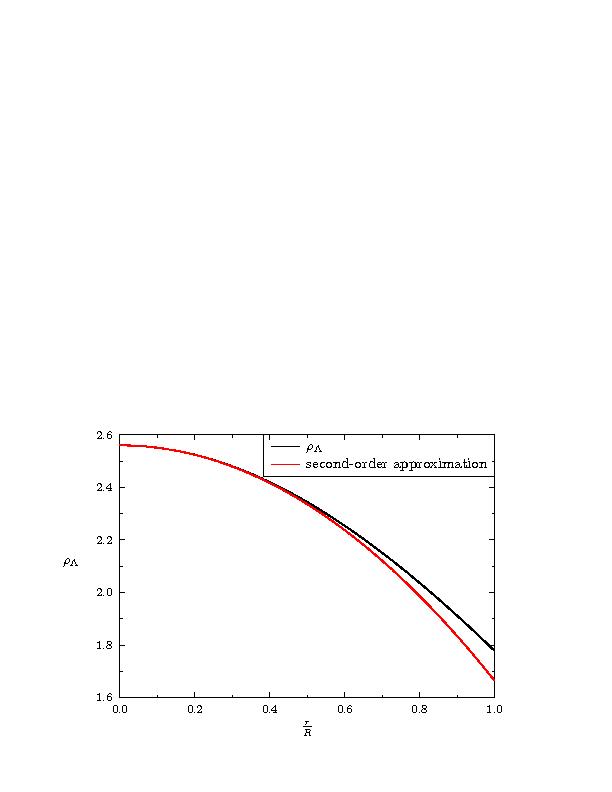}}}
\caption{ 
Dark energy density as a function of $\frac{r}{R}$, where $R$ is the radius of the star.
The upper curve depicts the result of the monopole approximation and the lower curve is an approximation
for the upper one. 
The figure is taken from \cite{caspar-de}.
\label{caspar-de}
}
\end{figure}

In \cite{caspar-de} calculations in one-loop order, using the monopole approximation \cite{caspar-de}, 
were calculated with the intention to
derive the coupling between the dark energy and matter density. In Fig. \ref{caspar-de} the lower curve shows the result of these calculations and the upper curve the approximation in terms of a polynomial, used in the final calculations.

\begin{figure}[t]
\rotatebox{0}{\resizebox{400pt}{300pt}{\includegraphics[width=0.5\textwidth]{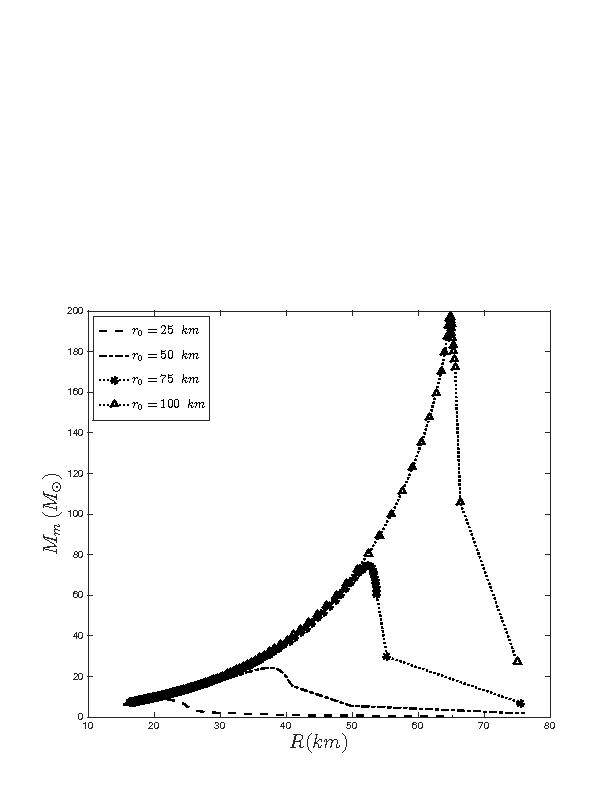}}}
\caption{ \label{caspar} The mass of a star as a function on its radius $R$. With the modified coupling 
of the dark-energy to the mass density the maximal mass possible is now about 200.
The figure is taken from \cite{caspar-de}.
}
\end{figure}

Finally, in Fig. \ref{caspar} the mass of the star versus its radius $R$ is depicted. As can be noted, now stars with up to 200
solar masses are possible. Higher masses cannot be obtained due to the limits the model of \cite{dexheimer} reaches.

This model also suffers from the approximations made and a complete description cannot be given. Nevertheless, that now
stars with up to 200 solar masses can be stabilized shows that the inclusion of dark-energy in massive stars may lead to
stable stars of any mass! (Though, only within a phenomenological model.)

\section{Conclusions}
\label{concl}

A report on the recent advances of the pseudo-complex General Relativity (pc-GR) was presented. The
theory predicts a non-zero energy-momentum tensor on the right hand side of the Einstein equation.
The new contribution is related to vacuum fluctuations, but due to a missing quantized theory of gravitation
one recurs to a phenomenological ansatz . Calculations in one-loop order, with
a constant back-ground metric, shows that the dark energy density has to increase toward smaller $r$.

Consequences of the theory were presented; i) The appearance of a dark ring followed by a bright one in accretion
discs around black holes, ii) a new interpretation of the source of the first gravitational event observed,
iii) possible outcomes of the future evolving universe and iv) attempts to
stabilize stars with large masses.

The only robust prediction is the structure in the emission profile of an accretion disc.

\section{Acknowledgment}
\label{secack}
{\it Peter O. Hess} acknowledges the financial support from DGAPA-PAPIIT (IN100418).
Very helpful discussions with T. Boller (Max-Planck Institure for Extraterrestial Physics, Garching, Germany) and
T. Sch\"onenbach are also acknowledged..

\end{document}